\documentclass[aps,prb,reprint,superscriptaddress]{revtex4-1}
\usepackage{amsmath}
\usepackage{multirow}
\usepackage{graphicx}
\usepackage{epstopdf}
\usepackage{tabularx, booktabs}
\newcolumntype{Y}{>{\centering\arraybackslash}X}
\usepackage[usenames,dvipsnames]{color}
\usepackage[caption=false,position=b,singlelinecheck=off,font=normalsize,labelfont=bf,justification=justified]{subfig}
\usepackage{hyperref}

\hypersetup{
    colorlinks=true,
    linkcolor=black,
    filecolor=black,      
    urlcolor=blue,
    citecolor=black,
}
\graphicspath{ {Figures_PDF/} }

\begin{document}

\title{Anisotropic exchange and spin-wave damping in pure and electron-doped Sr$_2$IrO$_4$}

\author{D. Pincini}
\email[]{davide.pincini@diamond.ac.uk}
\affiliation{London Centre for Nanotechnology and Department of Physics and Astronomy, University College London, Gower Street, London WC1E6BT, UK}
\affiliation{Diamond Light Source Ltd., Diamond House, Harwell Science \& Innovation Campus, Didcot, Oxfordshire OX11 0DE, UK }

\author{J.G. Vale}
\email[]{j.vale@ucl.ac.uk}
\affiliation{London Centre for Nanotechnology and Department of Physics and Astronomy, University College London, Gower Street, London WC1E6BT, UK}
\affiliation{Laboratory for Quantum Magnetism, Ecole Polytechnique F\'{e}d\'{e}rale de Lausanne (EPFL), CH-1015 Lausanne, Switzerland}

\author{C. Donnerer}
\affiliation{London Centre for Nanotechnology and Department of Physics and Astronomy, University College London, Gower Street, London WC1E6BT, UK}

\author{A. de la Torre}
\affiliation{California Institute of Technology, 1200 East California Blvd, 91125 Pasadena CA, USA}
\affiliation{Institute for Quantum Information and Matter, California Institute of Technology, Pasadena, California 91125, USA}

\author{E.C. Hunter}
\altaffiliation[Permanent address: ]{Inorganic Chemistry, University of Oxford, South Parks Road, OX1 3QR, UK}
\affiliation{SUPA, School of Physics and Astronomy, and Centre for Science at Extreme Conditions, The University of Edinburgh, Mayfield Road, Edinburgh EH9 3JZ}

\author{R. Perry}
\affiliation{London Centre for Nanotechnology and Department of Physics and Astronomy, University College London, Gower Street, London WC1E6BT, UK}

\author{M. Moretti Sala}
\affiliation{European Synchrotron Radiation Facility, BP 220, F-38043 Grenoble Cedex, France}

\author{F. Baumberger}
\affiliation{Department of Quantum Matter Physics, University of Geneva, 24 Quai Ernest-Ansermet, 1211 Geneva 4, Switzerland}
\affiliation{Swiss Light Source, Paul Scherrer Institut, CH-5232 Villigen PSI, Switzerland}

\author{D.F. McMorrow}
\affiliation{London Centre for Nanotechnology and Department of Physics and Astronomy, University College London, Gower Street, London WC1E6BT, UK}

\date{\today}

\begin{abstract}
The collective magnetic excitations in the spin-orbit Mott insulator (Sr$_{1-x}$La$_x$)$_2$IrO$_4$ ($x=0,\,0.01,\,0.04,\, 0.1$) were investigated by means of resonant inelastic x-ray scattering. We report significant magnon energy gaps at both the crystallographic and antiferromagnetic zone centers at all doping levels, along with a remarkably pronounced momentum-dependent lifetime broadening. The spin-wave gap is accounted for by a significant anisotropy in the interactions between $J_\text{eff}=1/2$ isospins, thus marking the departure of Sr$_2$IrO$_4$ from the essentially isotropic Heisenberg model appropriate for the superconducting cuprates.
\end{abstract}

\maketitle

\section{Introduction}

The combination of strong spin-orbit coupling in the presence of significant electron correlations leads to radically new electronic and magnetic phases, the elucidation of which is the subject of intense experimental and theoretical efforts \cite{review}. In their landmark paper, \citet{jackeli2009mott} established how
for the case of a $J_{\text{eff}}=1/2$ ground state, relevant for octahedrally coordinated \mbox{Ir$^{4+}$-based} transition-metal oxides (TMOs), a unique balance arises between anisotropic and isotropic exchanges, enshrined in the Kitaev-Heisenberg model, which depends exquisitely on lattice topology. Therefore determining the nature of anisotropic interactions is central to the program of understanding the novel physics displayed by $5d$ (and, indeed, $4d$) TMOs.\par
\begin{figure*}[htp]
\centering
\includegraphics[width=\textwidth]{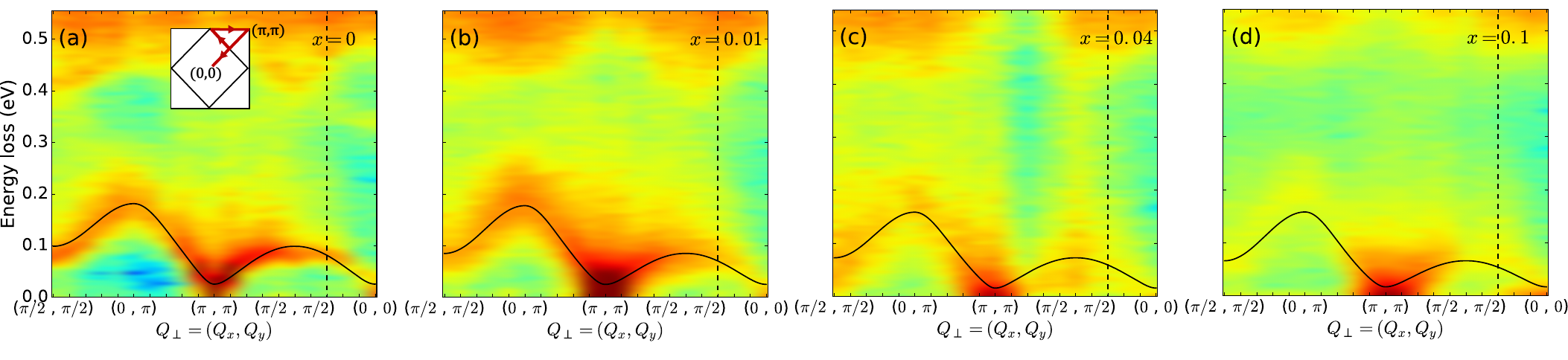}
\caption[]{(Color online) RIXS spectra along high-symmetry directions of the $(0,0,33)$ first BZ [see inset in (a)] for (a) $x=0$, (b) $x=0.01$, (c) $x=0.04$ and (d) $x=0.1$. The solid lines represent the fit to the 2DAH model discussed in the text (see Fig.~\ref{magnon_dispersion}). The spectra marked by the vertical dashed lines are plotted in Fig.~\ref{fit_detail}. The data from different samples were normalized to the spectral weight around $E_{\text{loss}}=1$~eV of the $(0,0)$ spectrum.}
\label{raw_data}
\end{figure*} 
Sr$_2$IrO$_4$ is of particular significance as the first spin-orbit Mott insulator to be identified \cite{PhysRevB.49.9198} and because of its electronic and structural similarities \cite{PhysRevB.49.9198,kim2014fermi,kim2016observation, de2015collapse, cao2016hallmarks} to the cuprate superconductor parent compound La$_2$CuO$_4$. This has led to the prediction of a superconducting state in doped Sr$_2$IrO$_4$ \cite{wang2011twisted,meng2014odd}. Analogous to the Cu$^{2+}$ ($S=1/2$) magnetic moments in La$_2$CuO$_4$, the spin-orbit entangled $J_{\text{eff}}=1/2$ \cite{kim2009phase,kim2008novel} \textit{isospins} of the Ir$^{4+}$ ions in Sr$_2$IrO$_4$ order at low temperature in a two-dimensional (2D) square lattice of antiferromagnetically coupled moments confined in the IrO$_2$ planes of the tetragonal crystal structure \cite{boseggia2013robustness,kim2009phase}. Upon doping Sr$_2$IrO$_4$ with electrons, the long-range order is suppressed, while short-range magnetic correlations have been shown to persist up to about $6\%$ La substitution \cite{liu2016anisotropic,gretarsson2016persistent,chen2015influence}. A similar behavior is also encountered in the hole-doped cuprates \cite{dean2013persistence}, as expected given the opposite sign of the next-nearest-neighbor hopping amplitude \cite{wang2011twisted}. Collective magnetic excitations with vanishing energy gap and a qualitatively similar energy dispersion have been reported to date in the parent and doped compounds of both La$_2$CuO$_4$ \cite{coldea2001spin, dean2013persistence} and Sr$_2$IrO$_4$ \cite{kim2012magnetic,liu2016anisotropic,gretarsson2016persistent}, which were interpreted in both cases in terms of a standard isotropic Heisenberg Hamiltonian extended to include next-nearest-neighbor interactions.\par 
The experimental reports of a purely isotropic exchange model for Sr$_2$IrO$_4$ are, in fact, at variance with the detailed predictions of \citet{jackeli2009mott}, who argued that there should be significant departures from the rotationally invariant Heisenberg model when Hund's coupling ($J_H=0.45$~eV in the case of Ir$^{4+}$ \cite{igarashi2014analysis,PhysRevB.88.104406}) and the deviation from cubic symmetry are incorporated. In this scenario, collective magnetic excitations will acquire a finite energy gap at both the crystallographic and antiferromagnetic (AF) zone centers.
Evidence supporting the presence of anisotropic magnetic interactions in Sr$_2$IrO$_4$ has come from a number of sources, including a detailed study of the magnetic critical scattering \cite{vale2015importance} and the observation of a small, zero-wave-vector magnon energy gap in electron spin resonance ($0.83$~meV) \cite{PhysRevB.89.180401} and Raman spectroscopy ($1.38$~meV) \cite{PhysRevB.93.024405}. Nonetheless, all previous resonant inelastic x-ray scattering (RIXS) investigations on the magnetic excitation spectrum of the parent \cite{kim2012magnetic} and electron-doped compounds \cite{liu2016anisotropic,gretarsson2016persistent} have not explicitly reported the presence of a gap, nor have they discussed the role of the anisotropic terms in the interaction Hamiltonian.\par

In this paper we report on a comprehensive RIXS study of the collective magnetic excitations in both parent Sr$_2$IrO$_4$ and its electron-doped version (Sr$_{1-x}$La$_x$)$_2$IrO$_4$. In contrast to earlier studies, we perform a full line-shape analysis of the RIXS spectra, including, most importantly, the effect of the finite momentum $Q$ and energy resolution. The excitation spectrum is shown to be fully gapped at all wave vectors in the Brillouin zone (BZ) up to $x=0.1$, indicating the existence of anisotropic exchange interactions, and a previously unreported anisotropic damping away from the zone centers is revealed.\par

\section{Samples and experimental setup}

Single crystals of (Sr$_{1-x}$La$_x$)$_2$IrO$_4$ with varying La concentration [$x=0,\,0.01(1),\,0.04(1),\,0.10(1)$] were flux grown using standard methods and characterized by resistivity and susceptibility measurements as described in Ref.~\onlinecite{de2015collapse} for samples of the same batch. The doping level of each of the crystals was checked by means of energy-dispersive x-ray spectroscopy (EDX). 
Substitution of trivalent La for divalent Sr dopes the system with electrons ($2x\,e^-/ \text{Ir}$ atom) and suppresses the long-range magnetic order for $x>x_c=0.02(1)$ \cite{chen2015influence}, while short-range correlations persist in the basal plane of the crystal \cite{chen2015influence,gretarsson2016persistent,de2015collapse,supplemental_material}.\par
The RIXS measurements were performed at the ID20 beamline of the European Synchrotron Radiation Facility (Grenoble, France). The experiment was carried out in horizontal scattering geometry using a spherical ($R=2$~m) Si(844) diced analyzer with a $60$~mm mask and a Si(844) secondary monochromator. This resulted in an overall energy resolution of $\text{FWHM}=23.4\,(28.0)$~meV for the $x=0,\,0.01,\,0.04\,(0.1)$ measurements and an in-plane momentum resolution of $\Delta Q_{\perp}\approx0.18\,\text{\AA}^{-1}$ \cite{supplemental_material}. The samples were cooled down to $T=20$~K (below the N\'{e}el transition at $T_N\approx 230$~K found in undoped Sr$_2$IrO$_4$ \cite{vale2015importance}) by means of a He-flow cryostat. 
 \begin{figure*}[htp]
\centering
\includegraphics[width=\textwidth]{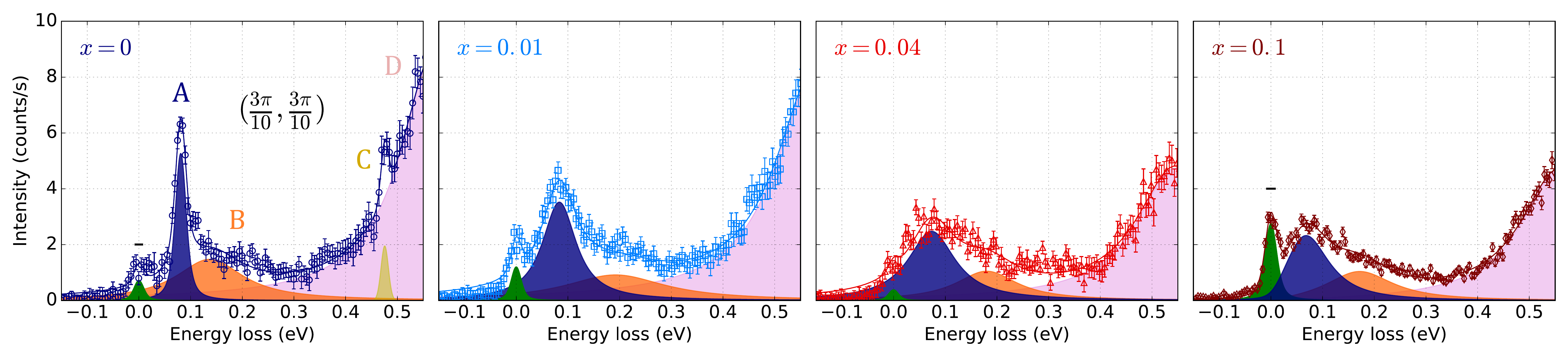}
\caption[]{(Color online) Fit detail of representative RIXS spectra for samples with different La contents. Open symbols and solid lines represent the measured intensity and the fit of the spectra, respectively. The Voigt functions used to fit the elastic line and the various inelastic contributions (discussed in the text) are displayed through the shaded regions. The horizontal lines in the $x=0$ (same as $x=0.01,\,0.04$) and $x=0.1$ panels represent the Gaussian width of the resolution function (fixed during the fit). The spectra from different samples were normalized to the spectral weight around $E_{\text{loss}}=1$~eV of the $(0,0)$ spectrum.}
\label{fit_detail}
\end{figure*} 
The in-plane momentum transfer values $\mathbf{Q}_{\perp}=(Q_x\,,\,Q_y)$ reported in this paper are quoted in units of $1/a$, where $a=3.89\,\text{\AA}$ is the in-plane lattice constant of the undistorted $I4/mmm$ unit cell. The out-of-plane component was kept fixed to $L=33$ for all the spectra ($L$ is the out-of-plane Miller index). The only exception is represented by the $(0,0)$ spectrum in the $x=0$ sample: this was measured for $L=32.85$ to minimize the strong elastic signal arising from the ordered magnetic structure.\par

\section{Results}
\subsection{Spin-wave excitation spectrum}
RIXS spectra were collected keeping the incident energy fixed to the Ir $L_3$ absorption edge and measuring the energy of the scattered photons in the energy loss range $E_{\text{loss}}=-0.2-0.6$~eV. For each value $x$ of La content, several spectra were collected for different values of $\mathbf{Q}_{\perp}$ along high-symmetry directions of the $(0,0,33)$ first BZ ($2\theta\approx90^\circ$). These are plotted in the intensity maps of Figs.~\ref{raw_data}(a)-\ref{raw_data}(d). As first reported by \citet{kim2012magnetic}, the parent compound data show a collective magnetic excitation dispersing from the AF zone center $(\pi,\pi)$ and extending up to about $0.2$~eV. In agreement with earlier studies \cite{gretarsson2016persistent, liu2016anisotropic}, damped magnetic excitations with a similar in-plane dispersion survive in the doped compounds deep into the metallic phase, where the long-range magnetic order is suppressed \cite{supplemental_material}. In particular, the magnons in our heavily doped ($x=0.1$) Sr$_2$IrO$_4$ sample still reflect the persistence of commensurate short-range order, in contrast to hole-doped La$_2$CuO$_4$ \cite{supplemental_material,hourglass1,hourglass2,PhysRevLett.67.1791}.\par
\begin{figure}[htp]
\centering
\includegraphics[width=\columnwidth]{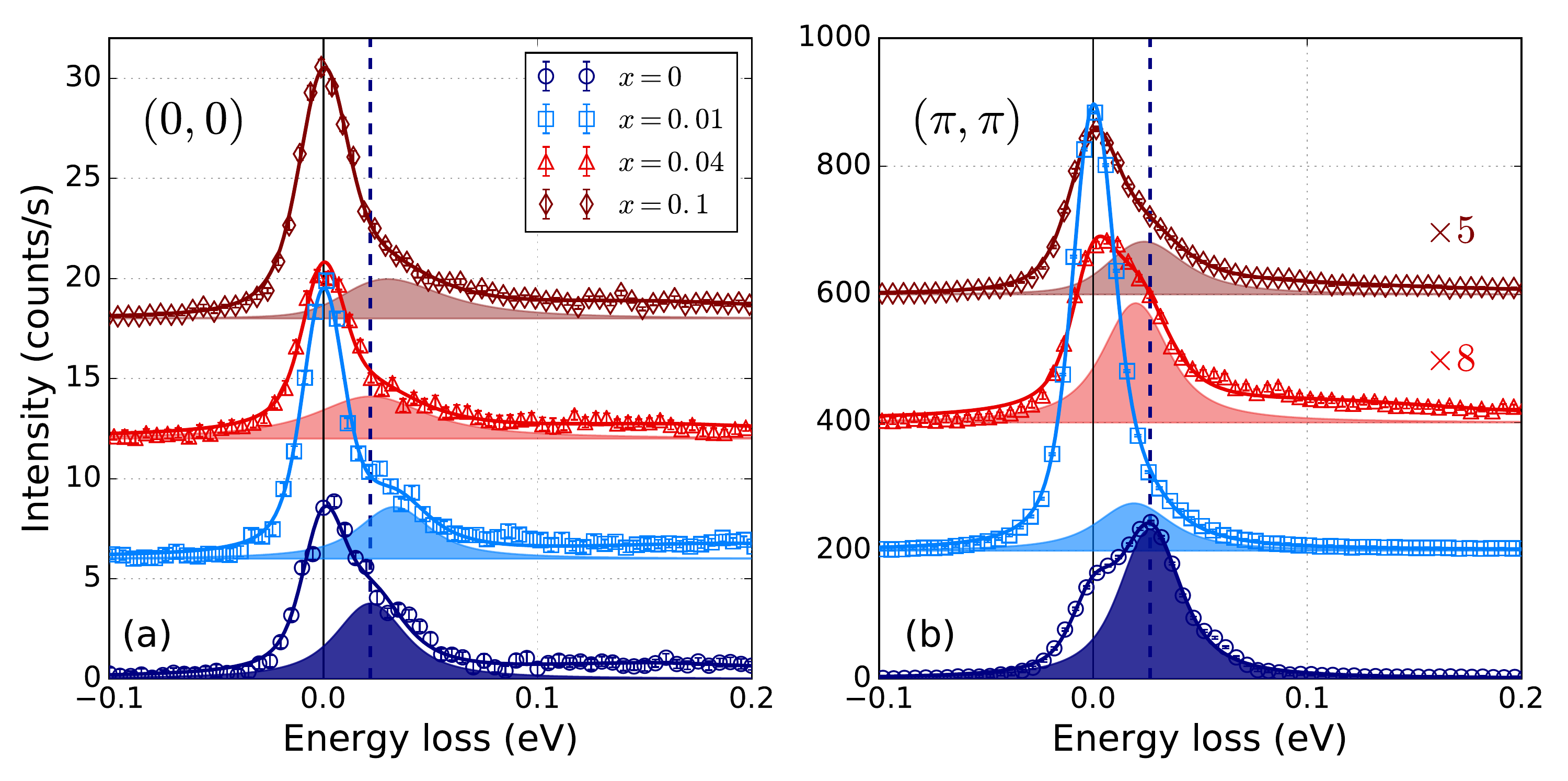}
\caption[]{(Color online) Doping level dependence of the low-energy region of the RIXS spectra at (a) $(0,0)$ and (b) $(\pi,\pi)$. The shaded regions show the fitted single-magnon peak, highlighting the presence of a finite spin-wave gap. The vertical dashed lines mark the value of the magnon energy in the parent compound $(x=0)$. The fitted magnon energies (and the corresponding $Q$-resolution-corrected values \cite{supplemental_material}) are reported in Table~\ref{table}.}
\label{magnon_detail}
\end{figure}
\begin{figure*}[htp]
\centering
\includegraphics[width=\textwidth]{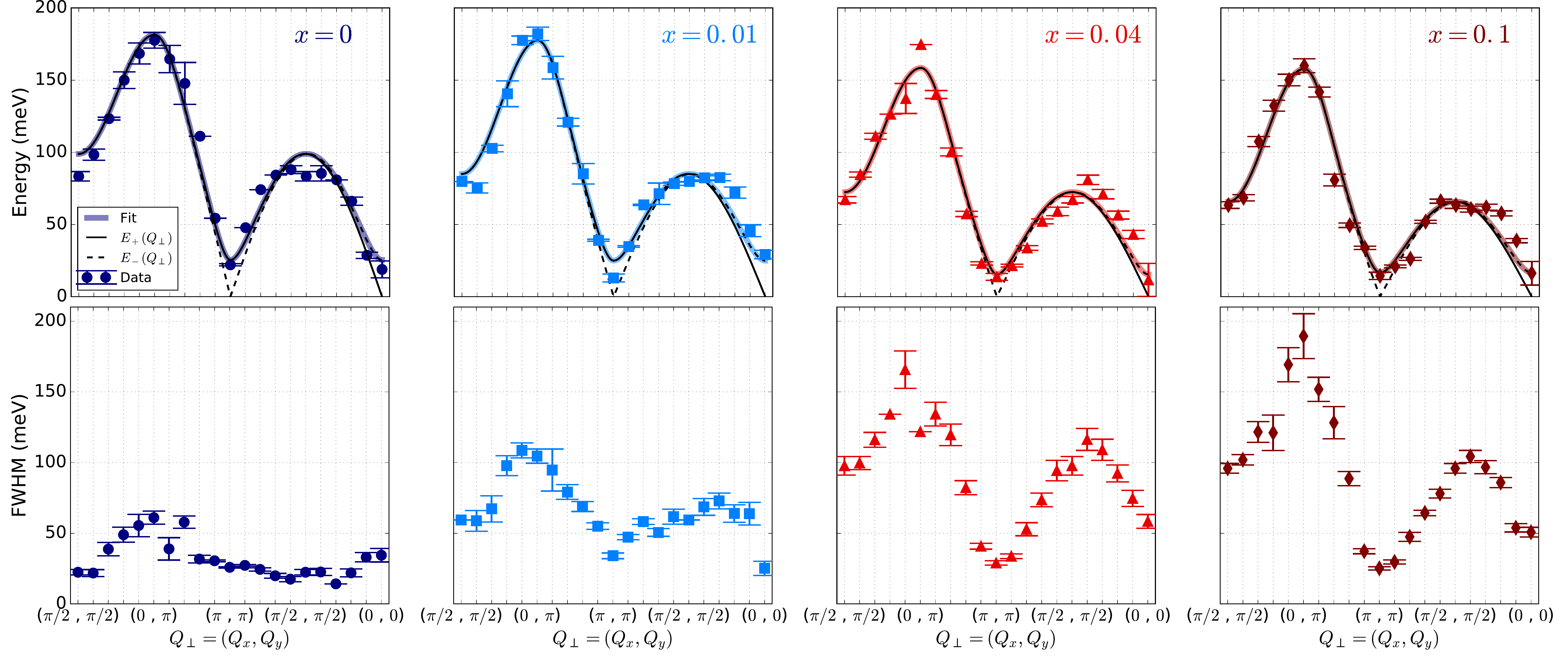}
\caption[]{(Color online) $Q$-resolution-corrected \cite{supplemental_material} energy dispersion (top) and Lorentzian lifetime broadening (bottom) of the single magnon for samples with different La contents. The errorbars represent $1\sigma$ confidence intervals [apart from the energy at $(0,0)$ and $(\pi,\pi)$, where the $2\sigma$ interval is reported instead]. The thick solid lines in color (top) show the fit to the 2DAH model as described in the text (best-fit values reported in Table~\ref{table}), while the thin solid and dashed black lines represent the two modes calculated for the best-fit parameters.}
\label{magnon_dispersion}
\end{figure*}
A quantitative analysis of the spin-wave spectrum was achieved by fitting the RIXS data by a sum of an elastic line and the following inelastic features (Fig.~\ref{fit_detail}): (A) single-magnon excitation, (B) a multimagnon continuum and (C) and (D) intra-$t_{2g}$ excitations \cite{kim2012magnetic,kim2014excitonic}. Each feature was modeled by a Voigt profile, with the width of the Gaussian component constrained to the experimental energy resolution: this allows the extraction of both the energy and the intrinsic Lorentzian lifetime broadening of the excitations~\cite{supplemental_material}.\par
One of the main features emerging from our data is the presence of a finite energy gap, which appears relatively robust with La doping. This is evident from the low-energy detail of the  spectra collected at the crystallographic and AF zone centers shown in Figs.~\ref{magnon_detail}(a) and \ref{magnon_detail}(b), respectively. Simple inspection reveals a separate energy-loss peak partially overlapping with the elastic line. Based on its energy and momentum dependence, we argue that it corresponds to a gapped spin-wave excitation. We note that the finite $Q$ resolution of the spectrometer can generally lead to an artificial gap at minima of the dispersion. However, the measured gap values at all doping levels (Table~\ref{table}) consistently exceed the artificial gaps simulated for the case of gapless excitations \cite{supplemental_material}.
This result is robust against both the statistical ($2\sigma$ confidence interval) and the estimated systematic error \cite{supplemental_material}: the presence of gapped magnons is thus to be considered an intrinsic property of the excitation spectrum in (Sr$_{1-x}$La$_x$)$_2$IrO$_4$. As shown in the Supplemental Material \cite{supplemental_material}, the impact of the $Q$ resolution can be factored out from the measured energy values, leading to an average gap of 19(9) and 16(4)~meV at $(0,0)$ and $(\pi,\pi)$, respectively, with no systematic doping dependence (see Table~\ref{table}). We note that recent electron spin resonance \cite{PhysRevB.89.180401} and Raman spectroscopy \cite{PhysRevB.93.024405} studies reported a smaller gap of, respectively, $0.83$ and $1.38$~meV at the crystallographic zone center. The origin of the discrepancy with the above analysis remains an open issue. Nonetheless, our value appears to be roughly consistent with a previous estimate in the undoped compound \cite{vale2015importance}.\par
 
\bgroup
\def\arraystretch{1.1}
\begin{table*}[htp]
\centering
\resizebox{\textwidth}{!}{

\begin{tabularx}{\textwidth}{c *{9}{Y}}
\hline
\multirow{2}{*}{$\mathbf{x}$} & \multirow{2}{*}{$\mathbf{\tilde{J}}$ \textbf{(meV)}} & \multirow{2}{*}{$\mathbf{J_2}$ \textbf{(meV)}} & \multirow{2}{*}{$\mathbf{J_3}$ \textbf{(meV)}} & \multirow{2}{*}{$\boldsymbol\Delta_\lambda$} & \multicolumn{2}{c}{\textbf{Measured gap (meV)}} & \multicolumn{2}{c}{\textbf{Corrected values (meV)}} \\
                              &                                                        &                                                  &                                                  &                                              & $\boldsymbol(\boldsymbol 0,\boldsymbol 0\boldsymbol)$          & $\boldsymbol(\boldsymbol\pi,\boldsymbol\pi\boldsymbol)$          & $\boldsymbol(\boldsymbol 0,\boldsymbol 0\boldsymbol)$                  & $\boldsymbol(\boldsymbol\pi,\boldsymbol\pi\boldsymbol)$                 \\ \hline
0                             & 57(1)                                                  & -16.5(4)                                         & 12.4(3)                                          & 0.05(2)                                      & 24(5)            & 27(4)              & 19(7)                    & 22(4)                        \\ \hline
0.01                          & 51(1)                                                  & -18.5(4)                                         & 13.9(3)                                          & 0.06(2)                                      & 33(5)            & 19(4)                & 29(7)                    & 13(4)                        \\ \hline
0.04                          & 44.5(5)                                                & -17.2(2)                                         & 12.9(1)                                          & 0.03(2)                                      & 22(6)            & 20(4)                & 12(11)                    & 14(4)                        \\ \hline
0.1                           & 41.7(5)                                                & -18.4(2)                                         & 13.8(2)                                          & 0.04(2)                                      & 23(6)            & 20(4)                & 16(8)                    & 14(4)                        \\ \hline    
\end{tabularx}
}
\caption{Equation~(\ref{eq:hamiltonian}) best-fit parameters as extracted from the $Q$-resolution-corrected single-magnon dispersion (Fig.~\ref{magnon_dispersion}) and measured and $Q$-resolution-corrected \cite{supplemental_material} energy gap values for the different doping concentrations. $\lvert J_2 \rvert/J_3$ was kept fixed to $1.33$ for all samples. The values in parentheses represent $1\sigma$ confidence intervals for the best-fit parameters, while they correspond to the maximum between the statistical $2\sigma$ confidence interval and the estimated systematic error \cite{supplemental_material} for the gap values.}
\label{table}
\end{table*}
\egroup 
 
The full energy dispersion (corrected for the finite $Q$ resolution \cite{supplemental_material}) of the single magnon for the different doping levels across the BZ is summarized in Fig.~\ref{magnon_dispersion} along with the corresponding Lorentzian FWHM. Besides displaying a finite energy gap, Fig.~\ref{magnon_dispersion} reveals that the magnon peak at $(0,0)$ and $(\pi,\pi)$ does not considerably broaden as the dopant concentration is increased (the $\text{FWHM}$ increases by only $60\%$ going from the parent to the heavily doped $x=0.1$ sample). On the other hand, a remarkably pronounced anisotropic broadening occurs away from the zone centers. The largest effect is seen at the zone boundaries $(0,\pi)$ and $(\pi/2,\pi/2)$, where the $\text{FWHM}$ increases by a factor of about $3$ and $4$, respectively, going from $x=0$ to $x=0.1$. Here, heavily damped magnetic excitations (paramagnons) are thus present.\par

\subsection{Magnetic Hamiltonian}

The spin-wave gap arises as a result of a significant easy-plane anisotropy in the exchange interaction between $J_{\text{eff}}=1/2$ isospins. In Ref.~\onlinecite{vale2015importance}, the magnetic critical scattering and RIXS data were found to be correctly described by the following two-dimensional anisotropic Heisenberg (2DAH) Hamiltonian:
\begin{align}
H=&\sum_{\langle i,j \rangle}\tilde{J}\left[S_i^xS_j^x+S_i^yS_j^y+(1-\Delta_\lambda)S_i^zS_j^z   \right]   \nonumber\\ + &\sum_{\langle\langle i,j \rangle\rangle}J_2\vec{S}_i\cdot\vec{S}_j+\sum_{\langle\langle\langle i,j \rangle\rangle\rangle}\,J_3\vec{S}_i\cdot\vec{S}_j,
\label{eq:hamiltonian}
\end{align}
where $\tilde{J}=J_1/(1-\Delta_\lambda)$ is an effective nearest-neighbor (NN) exchange integral depending on the in-plane anisotropy parameter $0\leq\Delta_\lambda\leq 1$ \cite{Keimer1993} and $J_2$ and $J_3$ model the next-NN.~and third-NN~exchange interactions, respectively. In the framework of linear spin-wave theory, Eq.~(\ref{eq:hamiltonian}) gives rise to two momentum-dependent magnetic modes $E_{\pm}(\mathbf{Q}_{\perp})$ \cite{supplemental_material}. For $\Delta_\lambda\neq0$, the latter are nondegenerate and display a finite energy gap at the crystallographic ($E_-$) and AF ($E_+$) zone centers (see Fig.~\ref{magnon_dispersion}). The two magnetic modes $E_{\pm}$ are not resolved in our measurements. The reason is that their energy is almost degenerate for most of the $\mathbf{Q}_{\perp}$ values explored, with a non-negligible splitting present only at the zone centers. Moreover, the gapless mode carries a vanishingly small spectral weight at $(0,0)$, while it is hidden by the elastic signal arising from a weak structural reflection \cite{PhysRevB.87.140406,PhysRevB.87.144405} at $(\pi,\pi)$. Only the gapped mode is thus expected to be visible in the RIXS spectra and to account for the observed gap. Following this reasoning, the measured dispersion was then fitted to (i) $E_+$ along the path $(\pi/2,\pi/2)\rightarrow(0,\pi)\rightarrow(\pi,\pi)\rightarrow(\pi/2,\pi/2)$ and (ii) $E_-$ along the path $(\pi/2,\pi/2)\rightarrow(0,0)$. The results are shown in Fig.~\ref{magnon_dispersion}, while the corresponding best-fit parameters are summarized in Table~\ref{table}.\par
The $Q$-resolution-corrected gap at all doping levels is correctly reproduced by a value of the easy-plane anisotropy $\Delta_\lambda$ in the range $0.03-0.06$ (Table~\ref{table}), in good agreement with previous theoretical predictions \cite{igarashi2014analysis,PhysRevB.88.104406,PhysRevB.92.235109} and experimental estimates \cite{vale2015importance} for the undoped compound. These values are significantly larger than the ones found in La$_2$CuO$_4$ \cite{Keimer1993}. Our results thus confirm the critical scattering data \cite{vale2015importance} and firmly establish the importance of easy-plane anisotropy in the low-energy Hamiltonian of Sr$_2$IrO$_4$. The expression of the anisotropic exchange of Eq.~(\ref{eq:hamiltonian}) is consistent with the dipolar-like term expected to arise as a result of finite Hund's coupling in the model by \citet{jackeli2009mott}. Considering the exchange parameters of Table~\ref{table}, the latter predicts values of the in-plane ($\Delta_{xy}\approx\Gamma_2=3$~meV) and out-of-plane ($\Delta_z\approx\sqrt{\tilde{J}\Gamma_1}=9$~meV) gaps, which, although smaller, are consistent in order of magnitude with the ones measured from the RIXS spectra. The anisotropy does not show any significant dependence on the La content within the experimental uncertainty, thus suggesting that it is robust with carrier doping. The impact of electron doping on the spin-wave energy dispersion is limited to a renormalization of the NN~exchange interaction: this decreases as $x$ is increased (Table~\ref{table}), in agreement with what was reported by \citet{gretarsson2016persistent}. The pronounced anisotropic broadening, however, suggests that the injection of free carriers causes a strong enhancement of the scattering processes along the zone boundary $(\pi/2,\pi/2)\rightarrow (0,\pi)$, which shortens the excitation lifetime with respect to the crystallographic and AF zone centers. Strikingly, a recent ARPES study of La-doped Sr$_2$IrO$_4$ \cite{de2015collapse} found coherent excitations in the form of Fermi arcs around $(\pi/2,\pi/2)$ coexisting with strongly interacting, pseudo-gapped states at $(\pi,0)$. Such a behavior might arise from strongly anisotropic coupling to magnetic fluctuations, qualitatively consistent with the anisotropic damping of spin waves reported here.\par

\section{Concluding remarks}

In conclusion, our RIXS investigation has revealed the presence of gapped collective magnetic excitations in the electron-doped spin-orbit Mott insulator (Sr$_{1-x}$La$_x$)$_2$IrO$_4$ up to $x=0.1$. The magnon is robust upon carrier doping at the crystallographic and AF zone center, while paramagnons exhibiting a pronounced anisotropic damping are found elsewhere in the BZ. Consistent with theoretical predictions \cite{jackeli2009mott}, the gap can be ascribed to a significant in-plane anisotropy in the interaction between the Ir$^{4+}$ $J_{\text{eff}}=1/2$ isospins that breaks the full rotational symmetry of the magnetic Hamiltonian. Despite apparent similarities with the superconducting cuprates, our results show that the spin-orbit entangled nature of the $J_{\text{eff}}=1/2$ ground state gives rise to magnetic interactions which differ significantly from the pure spin ones encountered in the cuprates. This will pave the way to a deeper understanding of the differences between the two classes of compounds in light of the long-sought-after superconductivity in iridate oxides.\par
\begin{acknowledgments}
The authors would like to thank M. Rossi (ID20, ESRF) and S. Boseggia (UCL) for the helpful discussions and the support provided during the data analysis. This work is supported by the UK Engineering and Physical Sciences Research Council (Grants No. EP/N027671/1 and No. EP/N034694/1) and by the Swiss National Science Foundation (Grant No. 200021- 146995).\par
D.~Pincini and J.G.~Vale contributed equally to this work.
\end{acknowledgments}

\bibliography{References}

\begin{thebibliography}{32}%
\makeatletter
\providecommand \@ifxundefined [1]{%
 \@ifx{#1\undefined}
}%
\providecommand \@ifnum [1]{%
 \ifnum #1\expandafter \@firstoftwo
 \else \expandafter \@secondoftwo
 \fi
}%
\providecommand \@ifx [1]{%
 \ifx #1\expandafter \@firstoftwo
 \else \expandafter \@secondoftwo
 \fi
}%
\providecommand \natexlab [1]{#1}%
\providecommand \enquote  [1]{``#1''}%
\providecommand \bibnamefont  [1]{#1}%
\providecommand \bibfnamefont [1]{#1}%
\providecommand \citenamefont [1]{#1}%
\providecommand \href@noop [0]{\@secondoftwo}%
\providecommand \href [0]{\begingroup \@sanitize@url \@href}%
\providecommand \@href[1]{\@@startlink{#1}\@@href}%
\providecommand \@@href[1]{\endgroup#1\@@endlink}%
\providecommand \@sanitize@url [0]{\catcode `\\12\catcode `\$12\catcode
  `\&12\catcode `\#12\catcode `\^12\catcode `\_12\catcode `\%12\relax}%
\providecommand \@@startlink[1]{}%
\providecommand \@@endlink[0]{}%
\providecommand \url  [0]{\begingroup\@sanitize@url \@url }%
\providecommand \@url [1]{\endgroup\@href {#1}{\urlprefix }}%
\providecommand \urlprefix  [0]{URL }%
\providecommand \Eprint [0]{\href }%
\providecommand \doibase [0]{http://dx.doi.org/}%
\providecommand \selectlanguage [0]{\@gobble}%
\providecommand \bibinfo  [0]{\@secondoftwo}%
\providecommand \bibfield  [0]{\@secondoftwo}%
\providecommand \translation [1]{[#1]}%
\providecommand \BibitemOpen [0]{}%
\providecommand \bibitemStop [0]{}%
\providecommand \bibitemNoStop [0]{.\EOS\space}%
\providecommand \EOS [0]{\spacefactor3000\relax}%
\providecommand \BibitemShut  [1]{\csname bibitem#1\endcsname}%
\let\auto@bib@innerbib\@empty
\bibitem [{\citenamefont {Witczak-Krempa}\ \emph {et~al.}(2014)\citenamefont
  {Witczak-Krempa}, \citenamefont {Chen}, \citenamefont {Kim},\ and\
  \citenamefont {Balents}}]{review}%
  \BibitemOpen
  \bibfield  {author} {\bibinfo {author} {\bibfnamefont {W.}~\bibnamefont
  {Witczak-Krempa}}, \bibinfo {author} {\bibfnamefont {G.}~\bibnamefont
  {Chen}}, \bibinfo {author} {\bibfnamefont {Y.~B.}\ \bibnamefont {Kim}}, \
  and\ \bibinfo {author} {\bibfnamefont {L.}~\bibnamefont {Balents}},\ }\href
  {\doibase 10.1146/annurev-conmatphys-020911-125138} {\bibfield  {journal}
  {\bibinfo  {journal} {Annu. Rev. Condens. Matter Phys.}\ }\textbf {\bibinfo
  {volume} {5}},\ \bibinfo {pages} {57} (\bibinfo {year} {2014})}\BibitemShut
  {NoStop}%
\bibitem [{\citenamefont {Jackeli}\ and\ \citenamefont
  {Khaliullin}(2009)}]{jackeli2009mott}%
  \BibitemOpen
  \bibfield  {author} {\bibinfo {author} {\bibfnamefont {G.}~\bibnamefont
  {Jackeli}}\ and\ \bibinfo {author} {\bibfnamefont {G.}~\bibnamefont
  {Khaliullin}},\ }\href {\doibase 10.1103/PhysRevLett.102.017205} {\bibfield
  {journal} {\bibinfo  {journal} {Phys. Rev. Lett.}\ }\textbf {\bibinfo
  {volume} {102}},\ \bibinfo {pages} {017205} (\bibinfo {year}
  {2009})}\BibitemShut {NoStop}%
\bibitem [{\citenamefont {Crawford}\ \emph {et~al.}(1994)\citenamefont
  {Crawford}, \citenamefont {Subramanian}, \citenamefont {Harlow},
  \citenamefont {Fernandez-Baca}, \citenamefont {Wang},\ and\ \citenamefont
  {Johnston}}]{PhysRevB.49.9198}%
  \BibitemOpen
  \bibfield  {author} {\bibinfo {author} {\bibfnamefont {M.~K.}\ \bibnamefont
  {Crawford}}, \bibinfo {author} {\bibfnamefont {M.~A.}\ \bibnamefont
  {Subramanian}}, \bibinfo {author} {\bibfnamefont {R.~L.}\ \bibnamefont
  {Harlow}}, \bibinfo {author} {\bibfnamefont {J.~A.}\ \bibnamefont
  {Fernandez-Baca}}, \bibinfo {author} {\bibfnamefont {Z.~R.}\ \bibnamefont
  {Wang}}, \ and\ \bibinfo {author} {\bibfnamefont {D.~C.}\ \bibnamefont
  {Johnston}},\ }\href {\doibase 10.1103/PhysRevB.49.9198} {\bibfield
  {journal} {\bibinfo  {journal} {Phys. Rev. B}\ }\textbf {\bibinfo {volume}
  {49}},\ \bibinfo {pages} {9198} (\bibinfo {year} {1994})}\BibitemShut
  {NoStop}%
\bibitem [{\citenamefont {Kim}\ \emph {et~al.}(2014{\natexlab{a}})\citenamefont
  {Kim}, \citenamefont {Krupin}, \citenamefont {Denlinger}, \citenamefont
  {Bostwick}, \citenamefont {Rotenberg}, \citenamefont {Zhao}, \citenamefont
  {Mitchell}, \citenamefont {Allen},\ and\ \citenamefont {Kim}}]{kim2014fermi}%
  \BibitemOpen
  \bibfield  {author} {\bibinfo {author} {\bibfnamefont {Y.~K.}\ \bibnamefont
  {Kim}}, \bibinfo {author} {\bibfnamefont {O.}~\bibnamefont {Krupin}},
  \bibinfo {author} {\bibfnamefont {J.~D.}\ \bibnamefont {Denlinger}}, \bibinfo
  {author} {\bibfnamefont {A.}~\bibnamefont {Bostwick}}, \bibinfo {author}
  {\bibfnamefont {E.}~\bibnamefont {Rotenberg}}, \bibinfo {author}
  {\bibfnamefont {Q.}~\bibnamefont {Zhao}}, \bibinfo {author} {\bibfnamefont
  {J.~F.}\ \bibnamefont {Mitchell}}, \bibinfo {author} {\bibfnamefont {J.~W.}\
  \bibnamefont {Allen}}, \ and\ \bibinfo {author} {\bibfnamefont {B.~J.}\
  \bibnamefont {Kim}},\ }\href {\doibase 10.1126/science.1251151} {\bibfield
  {journal} {\bibinfo  {journal} {Science}\ }\textbf {\bibinfo {volume}
  {345}},\ \bibinfo {pages} {187} (\bibinfo {year}
  {2014}{\natexlab{a}})}\BibitemShut {NoStop}%
\bibitem [{\citenamefont {Kim}\ \emph {et~al.}(2016)\citenamefont {Kim},
  \citenamefont {Sung}, \citenamefont {Denlinger},\ and\ \citenamefont
  {Kim}}]{kim2016observation}%
  \BibitemOpen
  \bibfield  {author} {\bibinfo {author} {\bibfnamefont {Y.}~\bibnamefont
  {Kim}}, \bibinfo {author} {\bibfnamefont {N.}~\bibnamefont {Sung}}, \bibinfo
  {author} {\bibfnamefont {J.}~\bibnamefont {Denlinger}}, \ and\ \bibinfo
  {author} {\bibfnamefont {B.}~\bibnamefont {Kim}},\ }\href {\doibase
  10.1038/nphys3503} {\bibfield  {journal} {\bibinfo  {journal} {Nat. Phys.}\
  }\textbf {\bibinfo {volume} {12}},\ \bibinfo {pages} {37} (\bibinfo {year}
  {2016})}\BibitemShut {NoStop}%
\bibitem [{\citenamefont {de~la Torre}\ \emph {et~al.}(2015)\citenamefont
  {de~la Torre}, \citenamefont {McKeown~Walker}, \citenamefont {Bruno},
  \citenamefont {Ricc\'o}, \citenamefont {Wang}, \citenamefont
  {Gutierrez~Lezama}, \citenamefont {Scheerer}, \citenamefont {Giriat},
  \citenamefont {Jaccard}, \citenamefont {Berthod}, \citenamefont {Kim},
  \citenamefont {Hoesch}, \citenamefont {Hunter}, \citenamefont {Perry},
  \citenamefont {Tamai},\ and\ \citenamefont {Baumberger}}]{de2015collapse}%
  \BibitemOpen
  \bibfield  {author} {\bibinfo {author} {\bibfnamefont {A.}~\bibnamefont
  {de~la Torre}}, \bibinfo {author} {\bibfnamefont {S.}~\bibnamefont
  {McKeown~Walker}}, \bibinfo {author} {\bibfnamefont {F.~Y.}\ \bibnamefont
  {Bruno}}, \bibinfo {author} {\bibfnamefont {S.}~\bibnamefont {Ricc\'o}},
  \bibinfo {author} {\bibfnamefont {Z.}~\bibnamefont {Wang}}, \bibinfo {author}
  {\bibfnamefont {I.}~\bibnamefont {Gutierrez~Lezama}}, \bibinfo {author}
  {\bibfnamefont {G.}~\bibnamefont {Scheerer}}, \bibinfo {author}
  {\bibfnamefont {G.}~\bibnamefont {Giriat}}, \bibinfo {author} {\bibfnamefont
  {D.}~\bibnamefont {Jaccard}}, \bibinfo {author} {\bibfnamefont
  {C.}~\bibnamefont {Berthod}}, \bibinfo {author} {\bibfnamefont {T.~K.}\
  \bibnamefont {Kim}}, \bibinfo {author} {\bibfnamefont {M.}~\bibnamefont
  {Hoesch}}, \bibinfo {author} {\bibfnamefont {E.~C.}\ \bibnamefont {Hunter}},
  \bibinfo {author} {\bibfnamefont {R.~S.}\ \bibnamefont {Perry}}, \bibinfo
  {author} {\bibfnamefont {A.}~\bibnamefont {Tamai}}, \ and\ \bibinfo {author}
  {\bibfnamefont {F.}~\bibnamefont {Baumberger}},\ }\href {\doibase
  10.1103/PhysRevLett.115.176402} {\bibfield  {journal} {\bibinfo  {journal}
  {Phys. Rev. Lett.}\ }\textbf {\bibinfo {volume} {115}},\ \bibinfo {pages}
  {176402} (\bibinfo {year} {2015})}\BibitemShut {NoStop}%
\bibitem [{\citenamefont {Cao}\ \emph {et~al.}(2016)\citenamefont {Cao},
  \citenamefont {Wang}, \citenamefont {Waugh}, \citenamefont {Reber},
  \citenamefont {Li}, \citenamefont {Zhou}, \citenamefont {Parham},
  \citenamefont {Park}, \citenamefont {Plumb}, \citenamefont {Rotenberg} \emph
  {et~al.}}]{cao2016hallmarks}%
  \BibitemOpen
  \bibfield  {author} {\bibinfo {author} {\bibfnamefont {Y.}~\bibnamefont
  {Cao}}, \bibinfo {author} {\bibfnamefont {Q.}~\bibnamefont {Wang}}, \bibinfo
  {author} {\bibfnamefont {J.~A.}\ \bibnamefont {Waugh}}, \bibinfo {author}
  {\bibfnamefont {T.~J.}\ \bibnamefont {Reber}}, \bibinfo {author}
  {\bibfnamefont {H.}~\bibnamefont {Li}}, \bibinfo {author} {\bibfnamefont
  {X.}~\bibnamefont {Zhou}}, \bibinfo {author} {\bibfnamefont {S.}~\bibnamefont
  {Parham}}, \bibinfo {author} {\bibfnamefont {S.}~\bibnamefont {Park}},
  \bibinfo {author} {\bibfnamefont {N.~C.}\ \bibnamefont {Plumb}}, \bibinfo
  {author} {\bibfnamefont {E.}~\bibnamefont {Rotenberg}},  \emph {et~al.},\
  }\href {http://www.nature.com/articles/ncomms11367} {\bibfield  {journal}
  {\bibinfo  {journal} {Nat. Commun.}\ }\textbf {\bibinfo {volume} {7}},\
  \bibinfo {pages} {11367} (\bibinfo {year} {2016})}\BibitemShut {NoStop}%
\bibitem [{\citenamefont {Wang}\ and\ \citenamefont
  {Senthil}(2011)}]{wang2011twisted}%
  \BibitemOpen
  \bibfield  {author} {\bibinfo {author} {\bibfnamefont {F.}~\bibnamefont
  {Wang}}\ and\ \bibinfo {author} {\bibfnamefont {T.}~\bibnamefont {Senthil}},\
  }\href {\doibase 10.1103/PhysRevLett.106.136402} {\bibfield  {journal}
  {\bibinfo  {journal} {Phys. Rev. Lett.}\ }\textbf {\bibinfo {volume} {106}},\
  \bibinfo {pages} {136402} (\bibinfo {year} {2011})}\BibitemShut {NoStop}%
\bibitem [{\citenamefont {Meng}\ \emph {et~al.}(2014)\citenamefont {Meng},
  \citenamefont {Kim},\ and\ \citenamefont {Kee}}]{meng2014odd}%
  \BibitemOpen
  \bibfield  {author} {\bibinfo {author} {\bibfnamefont {Z.~Y.}\ \bibnamefont
  {Meng}}, \bibinfo {author} {\bibfnamefont {Y.~B.}\ \bibnamefont {Kim}}, \
  and\ \bibinfo {author} {\bibfnamefont {H.~Y.}\ \bibnamefont {Kee}},\ }\href
  {\doibase 10.1103/PhysRevLett.113.177003} {\bibfield  {journal} {\bibinfo
  {journal} {Phys. Rev. Lett.}\ }\textbf {\bibinfo {volume} {113}},\ \bibinfo
  {pages} {177003} (\bibinfo {year} {2014})}\BibitemShut {NoStop}%
\bibitem [{\citenamefont {Kim}\ \emph {et~al.}(2009)\citenamefont {Kim},
  \citenamefont {Ohsumi}, \citenamefont {Komesu}, \citenamefont {Sakai},
  \citenamefont {Morita}, \citenamefont {Takagi},\ and\ \citenamefont
  {Arima}}]{kim2009phase}%
  \BibitemOpen
  \bibfield  {author} {\bibinfo {author} {\bibfnamefont {B.}~\bibnamefont
  {Kim}}, \bibinfo {author} {\bibfnamefont {H.}~\bibnamefont {Ohsumi}},
  \bibinfo {author} {\bibfnamefont {T.}~\bibnamefont {Komesu}}, \bibinfo
  {author} {\bibfnamefont {S.}~\bibnamefont {Sakai}}, \bibinfo {author}
  {\bibfnamefont {T.}~\bibnamefont {Morita}}, \bibinfo {author} {\bibfnamefont
  {H.}~\bibnamefont {Takagi}}, \ and\ \bibinfo {author} {\bibfnamefont
  {T.}~\bibnamefont {Arima}},\ }\href {\doibase 10.1126/science.1167106}
  {\bibfield  {journal} {\bibinfo  {journal} {Science}\ }\textbf {\bibinfo
  {volume} {323}},\ \bibinfo {pages} {1329} (\bibinfo {year}
  {2009})}\BibitemShut {NoStop}%
\bibitem [{\citenamefont {Kim}\ \emph {et~al.}(2008)\citenamefont {Kim},
  \citenamefont {Jin}, \citenamefont {Moon}, \citenamefont {Kim}, \citenamefont
  {Park}, \citenamefont {Leem}, \citenamefont {Yu}, \citenamefont {Noh},
  \citenamefont {Kim}, \citenamefont {Oh} \emph {et~al.}}]{kim2008novel}%
  \BibitemOpen
  \bibfield  {author} {\bibinfo {author} {\bibfnamefont {B.}~\bibnamefont
  {Kim}}, \bibinfo {author} {\bibfnamefont {H.}~\bibnamefont {Jin}}, \bibinfo
  {author} {\bibfnamefont {S.}~\bibnamefont {Moon}}, \bibinfo {author}
  {\bibfnamefont {J.}~\bibnamefont {Kim}}, \bibinfo {author} {\bibfnamefont
  {B.-G.}\ \bibnamefont {Park}}, \bibinfo {author} {\bibfnamefont
  {C.}~\bibnamefont {Leem}}, \bibinfo {author} {\bibfnamefont {J.}~\bibnamefont
  {Yu}}, \bibinfo {author} {\bibfnamefont {T.}~\bibnamefont {Noh}}, \bibinfo
  {author} {\bibfnamefont {C.}~\bibnamefont {Kim}}, \bibinfo {author}
  {\bibfnamefont {S.}~\bibnamefont {Oh}},  \emph {et~al.},\ }\href {\doibase
  10.1103/PhysRevLett.101.076402} {\bibfield  {journal} {\bibinfo  {journal}
  {Phys. Rev. Lett.}\ }\textbf {\bibinfo {volume} {101}},\ \bibinfo {pages}
  {076402} (\bibinfo {year} {2008})}\BibitemShut {NoStop}%
\bibitem [{\citenamefont {Boseggia}\ \emph {et~al.}(2013)\citenamefont
  {Boseggia}, \citenamefont {Springell}, \citenamefont {Walker}, \citenamefont
  {R{\o}nnow}, \citenamefont {R{\"u}egg}, \citenamefont {Okabe}, \citenamefont
  {Isobe}, \citenamefont {Perry}, \citenamefont {Collins},\ and\ \citenamefont
  {McMorrow}}]{boseggia2013robustness}%
  \BibitemOpen
  \bibfield  {author} {\bibinfo {author} {\bibfnamefont {S.}~\bibnamefont
  {Boseggia}}, \bibinfo {author} {\bibfnamefont {R.}~\bibnamefont {Springell}},
  \bibinfo {author} {\bibfnamefont {H.~C.}\ \bibnamefont {Walker}}, \bibinfo
  {author} {\bibfnamefont {H.~M.}\ \bibnamefont {R{\o}nnow}}, \bibinfo {author}
  {\bibfnamefont {C.}~\bibnamefont {R{\"u}egg}}, \bibinfo {author}
  {\bibfnamefont {H.}~\bibnamefont {Okabe}}, \bibinfo {author} {\bibfnamefont
  {M.}~\bibnamefont {Isobe}}, \bibinfo {author} {\bibfnamefont {R.~S.}\
  \bibnamefont {Perry}}, \bibinfo {author} {\bibfnamefont {S.~P.}\ \bibnamefont
  {Collins}}, \ and\ \bibinfo {author} {\bibfnamefont {D.~F.}\ \bibnamefont
  {McMorrow}},\ }\href {\doibase 10.1103/PhysRevLett.110.117207} {\bibfield
  {journal} {\bibinfo  {journal} {Phys. Rev. Lett.}\ }\textbf {\bibinfo
  {volume} {110}},\ \bibinfo {pages} {117207} (\bibinfo {year}
  {2013})}\BibitemShut {NoStop}%
\bibitem [{\citenamefont {Liu}\ \emph {et~al.}(2016)\citenamefont {Liu},
  \citenamefont {Dean}, \citenamefont {Meng}, \citenamefont {Upton},
  \citenamefont {Qi}, \citenamefont {Gog}, \citenamefont {Cao}, \citenamefont
  {Lin}, \citenamefont {Meyers}, \citenamefont {Ding} \emph
  {et~al.}}]{liu2016anisotropic}%
  \BibitemOpen
  \bibfield  {author} {\bibinfo {author} {\bibfnamefont {X.}~\bibnamefont
  {Liu}}, \bibinfo {author} {\bibfnamefont {M.}~\bibnamefont {Dean}}, \bibinfo
  {author} {\bibfnamefont {Z.}~\bibnamefont {Meng}}, \bibinfo {author}
  {\bibfnamefont {M.}~\bibnamefont {Upton}}, \bibinfo {author} {\bibfnamefont
  {T.}~\bibnamefont {Qi}}, \bibinfo {author} {\bibfnamefont {T.}~\bibnamefont
  {Gog}}, \bibinfo {author} {\bibfnamefont {Y.}~\bibnamefont {Cao}}, \bibinfo
  {author} {\bibfnamefont {J.}~\bibnamefont {Lin}}, \bibinfo {author}
  {\bibfnamefont {D.}~\bibnamefont {Meyers}}, \bibinfo {author} {\bibfnamefont
  {H.}~\bibnamefont {Ding}},  \emph {et~al.},\ }\href {\doibase
  10.1103/PhysRevB.93.241102} {\bibfield  {journal} {\bibinfo  {journal} {Phys.
  Rev. B}\ }\textbf {\bibinfo {volume} {93}},\ \bibinfo {pages} {241102}
  (\bibinfo {year} {2016})}\BibitemShut {NoStop}%
\bibitem [{\citenamefont {Gretarsson}\ \emph {et~al.}(2016)\citenamefont
  {Gretarsson}, \citenamefont {Sung}, \citenamefont {Porras}, \citenamefont
  {Bertinshaw}, \citenamefont {Dietl}, \citenamefont {Bruin}, \citenamefont
  {Bangura}, \citenamefont {Kim}, \citenamefont {Dinnebier}, \citenamefont
  {Kim} \emph {et~al.}}]{gretarsson2016persistent}%
  \BibitemOpen
  \bibfield  {author} {\bibinfo {author} {\bibfnamefont {H.}~\bibnamefont
  {Gretarsson}}, \bibinfo {author} {\bibfnamefont {N.}~\bibnamefont {Sung}},
  \bibinfo {author} {\bibfnamefont {J.}~\bibnamefont {Porras}}, \bibinfo
  {author} {\bibfnamefont {J.}~\bibnamefont {Bertinshaw}}, \bibinfo {author}
  {\bibfnamefont {C.}~\bibnamefont {Dietl}}, \bibinfo {author} {\bibfnamefont
  {J.~A.}\ \bibnamefont {Bruin}}, \bibinfo {author} {\bibfnamefont
  {A.}~\bibnamefont {Bangura}}, \bibinfo {author} {\bibfnamefont
  {Y.}~\bibnamefont {Kim}}, \bibinfo {author} {\bibfnamefont {R.}~\bibnamefont
  {Dinnebier}}, \bibinfo {author} {\bibfnamefont {J.}~\bibnamefont {Kim}},
  \emph {et~al.},\ }\href {\doibase 10.1103/PhysRevLett.117.107001} {\bibfield
  {journal} {\bibinfo  {journal} {Phys. Rev. Lett.}\ }\textbf {\bibinfo
  {volume} {117}},\ \bibinfo {pages} {107001} (\bibinfo {year}
  {2016})}\BibitemShut {NoStop}%
\bibitem [{\citenamefont {Chen}\ \emph {et~al.}(2015)\citenamefont {Chen},
  \citenamefont {Hogan}, \citenamefont {Walkup}, \citenamefont {Zhou},
  \citenamefont {Pokharel}, \citenamefont {Yao}, \citenamefont {Tian},
  \citenamefont {Ward}, \citenamefont {Zhao}, \citenamefont {Parshall} \emph
  {et~al.}}]{chen2015influence}%
  \BibitemOpen
  \bibfield  {author} {\bibinfo {author} {\bibfnamefont {X.}~\bibnamefont
  {Chen}}, \bibinfo {author} {\bibfnamefont {T.}~\bibnamefont {Hogan}},
  \bibinfo {author} {\bibfnamefont {D.}~\bibnamefont {Walkup}}, \bibinfo
  {author} {\bibfnamefont {W.}~\bibnamefont {Zhou}}, \bibinfo {author}
  {\bibfnamefont {M.}~\bibnamefont {Pokharel}}, \bibinfo {author}
  {\bibfnamefont {M.}~\bibnamefont {Yao}}, \bibinfo {author} {\bibfnamefont
  {W.}~\bibnamefont {Tian}}, \bibinfo {author} {\bibfnamefont {T.~Z.}\
  \bibnamefont {Ward}}, \bibinfo {author} {\bibfnamefont {Y.}~\bibnamefont
  {Zhao}}, \bibinfo {author} {\bibfnamefont {D.}~\bibnamefont {Parshall}},
  \emph {et~al.},\ }\href {\doibase 10.1103/PhysRevB.92.075125} {\bibfield
  {journal} {\bibinfo  {journal} {Phys. Rev. B}\ }\textbf {\bibinfo {volume}
  {92}},\ \bibinfo {pages} {075125} (\bibinfo {year} {2015})}\BibitemShut
  {NoStop}%
\bibitem [{\citenamefont {Dean}\ \emph {et~al.}(2013)\citenamefont {Dean},
  \citenamefont {Dellea}, \citenamefont {Springell}, \citenamefont
  {Yakhou-Harris}, \citenamefont {Kummer}, \citenamefont {Brookes},
  \citenamefont {Liu}, \citenamefont {Sun}, \citenamefont {Strle},
  \citenamefont {Schmitt} \emph {et~al.}}]{dean2013persistence}%
  \BibitemOpen
  \bibfield  {author} {\bibinfo {author} {\bibfnamefont {M.}~\bibnamefont
  {Dean}}, \bibinfo {author} {\bibfnamefont {G.}~\bibnamefont {Dellea}},
  \bibinfo {author} {\bibfnamefont {R.}~\bibnamefont {Springell}}, \bibinfo
  {author} {\bibfnamefont {F.}~\bibnamefont {Yakhou-Harris}}, \bibinfo {author}
  {\bibfnamefont {K.}~\bibnamefont {Kummer}}, \bibinfo {author} {\bibfnamefont
  {N.}~\bibnamefont {Brookes}}, \bibinfo {author} {\bibfnamefont
  {X.}~\bibnamefont {Liu}}, \bibinfo {author} {\bibfnamefont {Y.}~\bibnamefont
  {Sun}}, \bibinfo {author} {\bibfnamefont {J.}~\bibnamefont {Strle}}, \bibinfo
  {author} {\bibfnamefont {T.}~\bibnamefont {Schmitt}},  \emph {et~al.},\
  }\href {\doibase doi:10.1038/nmat3723} {\bibfield  {journal} {\bibinfo
  {journal} {Nat. Mater.}\ }\textbf {\bibinfo {volume} {12}},\ \bibinfo {pages}
  {1019} (\bibinfo {year} {2013})}\BibitemShut {NoStop}%
\bibitem [{\citenamefont {Coldea}\ \emph {et~al.}(2001)\citenamefont {Coldea},
  \citenamefont {Hayden}, \citenamefont {Aeppli}, \citenamefont {Perring},
  \citenamefont {Frost}, \citenamefont {Mason}, \citenamefont {Cheong},\ and\
  \citenamefont {Fisk}}]{coldea2001spin}%
  \BibitemOpen
  \bibfield  {author} {\bibinfo {author} {\bibfnamefont {R.}~\bibnamefont
  {Coldea}}, \bibinfo {author} {\bibfnamefont {S.~M.}\ \bibnamefont {Hayden}},
  \bibinfo {author} {\bibfnamefont {G.}~\bibnamefont {Aeppli}}, \bibinfo
  {author} {\bibfnamefont {T.~G.}\ \bibnamefont {Perring}}, \bibinfo {author}
  {\bibfnamefont {C.~D.}\ \bibnamefont {Frost}}, \bibinfo {author}
  {\bibfnamefont {T.~E.}\ \bibnamefont {Mason}}, \bibinfo {author}
  {\bibfnamefont {S.~W.}\ \bibnamefont {Cheong}}, \ and\ \bibinfo {author}
  {\bibfnamefont {Z.}~\bibnamefont {Fisk}},\ }\href {\doibase
  10.1103/PhysRevLett.86.5377} {\bibfield  {journal} {\bibinfo  {journal}
  {Phys. Rev. Lett.}\ }\textbf {\bibinfo {volume} {86}},\ \bibinfo {pages}
  {5377} (\bibinfo {year} {2001})}\BibitemShut {NoStop}%
\bibitem [{\citenamefont {Kim}\ \emph {et~al.}(2012)\citenamefont {Kim},
  \citenamefont {Casa}, \citenamefont {Upton}, \citenamefont {Gog},
  \citenamefont {Kim}, \citenamefont {Mitchell}, \citenamefont
  {Van~Veenendaal}, \citenamefont {Daghofer}, \citenamefont {van Den~Brink},
  \citenamefont {Khaliullin} \emph {et~al.}}]{kim2012magnetic}%
  \BibitemOpen
  \bibfield  {author} {\bibinfo {author} {\bibfnamefont {J.}~\bibnamefont
  {Kim}}, \bibinfo {author} {\bibfnamefont {D.}~\bibnamefont {Casa}}, \bibinfo
  {author} {\bibfnamefont {M.}~\bibnamefont {Upton}}, \bibinfo {author}
  {\bibfnamefont {T.}~\bibnamefont {Gog}}, \bibinfo {author} {\bibfnamefont
  {Y.}~\bibnamefont {Kim}}, \bibinfo {author} {\bibfnamefont {J.}~\bibnamefont
  {Mitchell}}, \bibinfo {author} {\bibfnamefont {M.}~\bibnamefont
  {Van~Veenendaal}}, \bibinfo {author} {\bibfnamefont {M.}~\bibnamefont
  {Daghofer}}, \bibinfo {author} {\bibfnamefont {J.}~\bibnamefont {van
  Den~Brink}}, \bibinfo {author} {\bibfnamefont {G.}~\bibnamefont
  {Khaliullin}},  \emph {et~al.},\ }\href {\doibase
  10.1103/PhysRevLett.108.177003} {\bibfield  {journal} {\bibinfo  {journal}
  {Phys. Rev. Lett.}\ }\textbf {\bibinfo {volume} {108}},\ \bibinfo {pages}
  {177003} (\bibinfo {year} {2012})}\BibitemShut {NoStop}%
\bibitem [{\citenamefont {Igarashi}\ and\ \citenamefont
  {Nagao}(2014)}]{igarashi2014analysis}%
  \BibitemOpen
  \bibfield  {author} {\bibinfo {author} {\bibfnamefont {J.~I.}\ \bibnamefont
  {Igarashi}}\ and\ \bibinfo {author} {\bibfnamefont {T.}~\bibnamefont
  {Nagao}},\ }\href {\doibase 10.1103/PhysRevB.90.064402} {\bibfield  {journal}
  {\bibinfo  {journal} {Phys. Rev. B}\ }\textbf {\bibinfo {volume} {90}},\
  \bibinfo {pages} {064402} (\bibinfo {year} {2014})}\BibitemShut {NoStop}%
\bibitem [{\citenamefont {Igarashi}\ and\ \citenamefont
  {Nagao}(2013)}]{PhysRevB.88.104406}%
  \BibitemOpen
  \bibfield  {author} {\bibinfo {author} {\bibfnamefont {J.~I.}\ \bibnamefont
  {Igarashi}}\ and\ \bibinfo {author} {\bibfnamefont {T.}~\bibnamefont
  {Nagao}},\ }\href {\doibase 10.1103/PhysRevB.88.104406} {\bibfield  {journal}
  {\bibinfo  {journal} {Phys. Rev. B}\ }\textbf {\bibinfo {volume} {88}},\
  \bibinfo {pages} {104406} (\bibinfo {year} {2013})}\BibitemShut {NoStop}%
\bibitem [{\citenamefont {Vale}\ \emph {et~al.}(2015)\citenamefont {Vale},
  \citenamefont {Boseggia}, \citenamefont {Walker}, \citenamefont {Springell},
  \citenamefont {Feng}, \citenamefont {Hunter}, \citenamefont {Perry},
  \citenamefont {Prabhakaran}, \citenamefont {Boothroyd}, \citenamefont
  {Collins} \emph {et~al.}}]{vale2015importance}%
  \BibitemOpen
  \bibfield  {author} {\bibinfo {author} {\bibfnamefont {J.}~\bibnamefont
  {Vale}}, \bibinfo {author} {\bibfnamefont {S.}~\bibnamefont {Boseggia}},
  \bibinfo {author} {\bibfnamefont {H.}~\bibnamefont {Walker}}, \bibinfo
  {author} {\bibfnamefont {R.}~\bibnamefont {Springell}}, \bibinfo {author}
  {\bibfnamefont {Z.}~\bibnamefont {Feng}}, \bibinfo {author} {\bibfnamefont
  {E.}~\bibnamefont {Hunter}}, \bibinfo {author} {\bibfnamefont
  {R.}~\bibnamefont {Perry}}, \bibinfo {author} {\bibfnamefont
  {D.}~\bibnamefont {Prabhakaran}}, \bibinfo {author} {\bibfnamefont
  {A.}~\bibnamefont {Boothroyd}}, \bibinfo {author} {\bibfnamefont
  {S.}~\bibnamefont {Collins}},  \emph {et~al.},\ }\href {\doibase
  10.1103/PhysRevB.92.020406} {\bibfield  {journal} {\bibinfo  {journal} {Phys.
  Rev. B}\ }\textbf {\bibinfo {volume} {92}},\ \bibinfo {pages} {020406}
  (\bibinfo {year} {2015})}\BibitemShut {NoStop}%
\bibitem [{\citenamefont {Bahr}\ \emph {et~al.}(2014)\citenamefont {Bahr},
  \citenamefont {Alfonsov}, \citenamefont {Jackeli}, \citenamefont
  {Khaliullin}, \citenamefont {Matsumoto}, \citenamefont {Takayama},
  \citenamefont {Takagi}, \citenamefont {B\"uchner},\ and\ \citenamefont
  {Kataev}}]{PhysRevB.89.180401}%
  \BibitemOpen
  \bibfield  {author} {\bibinfo {author} {\bibfnamefont {S.}~\bibnamefont
  {Bahr}}, \bibinfo {author} {\bibfnamefont {A.}~\bibnamefont {Alfonsov}},
  \bibinfo {author} {\bibfnamefont {G.}~\bibnamefont {Jackeli}}, \bibinfo
  {author} {\bibfnamefont {G.}~\bibnamefont {Khaliullin}}, \bibinfo {author}
  {\bibfnamefont {A.}~\bibnamefont {Matsumoto}}, \bibinfo {author}
  {\bibfnamefont {T.}~\bibnamefont {Takayama}}, \bibinfo {author}
  {\bibfnamefont {H.}~\bibnamefont {Takagi}}, \bibinfo {author} {\bibfnamefont
  {B.}~\bibnamefont {B\"uchner}}, \ and\ \bibinfo {author} {\bibfnamefont
  {V.}~\bibnamefont {Kataev}},\ }\href {\doibase 10.1103/PhysRevB.89.180401}
  {\bibfield  {journal} {\bibinfo  {journal} {Phys. Rev. B}\ }\textbf {\bibinfo
  {volume} {89}},\ \bibinfo {pages} {180401} (\bibinfo {year}
  {2014})}\BibitemShut {NoStop}%
\bibitem [{\citenamefont {Gim}\ \emph {et~al.}(2016)\citenamefont {Gim},
  \citenamefont {Sethi}, \citenamefont {Zhao}, \citenamefont {Mitchell},
  \citenamefont {Cao},\ and\ \citenamefont {Cooper}}]{PhysRevB.93.024405}%
  \BibitemOpen
  \bibfield  {author} {\bibinfo {author} {\bibfnamefont {Y.}~\bibnamefont
  {Gim}}, \bibinfo {author} {\bibfnamefont {A.}~\bibnamefont {Sethi}}, \bibinfo
  {author} {\bibfnamefont {Q.}~\bibnamefont {Zhao}}, \bibinfo {author}
  {\bibfnamefont {J.~F.}\ \bibnamefont {Mitchell}}, \bibinfo {author}
  {\bibfnamefont {G.}~\bibnamefont {Cao}}, \ and\ \bibinfo {author}
  {\bibfnamefont {S.~L.}\ \bibnamefont {Cooper}},\ }\href {\doibase
  10.1103/PhysRevB.93.024405} {\bibfield  {journal} {\bibinfo  {journal} {Phys.
  Rev. B}\ }\textbf {\bibinfo {volume} {93}},\ \bibinfo {pages} {024405}
  (\bibinfo {year} {2016})}\BibitemShut {NoStop}%
\bibitem [{sup()}]{supplemental_material}%
  \BibitemOpen
  \href@noop {} {}\bibinfo {howpublished} {Supplemental material}\BibitemShut
  {NoStop}%
\bibitem [{\citenamefont {Fujita}\ \emph {et~al.}(2012)\citenamefont {Fujita},
  \citenamefont {Hiraka}, \citenamefont {Matsuda}, \citenamefont {Matsuura},
  \citenamefont {Tranquada}, \citenamefont {Wakimoto}, \citenamefont {Xu},\
  and\ \citenamefont {Yamada}}]{hourglass1}%
  \BibitemOpen
  \bibfield  {author} {\bibinfo {author} {\bibfnamefont {M.}~\bibnamefont
  {Fujita}}, \bibinfo {author} {\bibfnamefont {H.}~\bibnamefont {Hiraka}},
  \bibinfo {author} {\bibfnamefont {M.}~\bibnamefont {Matsuda}}, \bibinfo
  {author} {\bibfnamefont {M.}~\bibnamefont {Matsuura}}, \bibinfo {author}
  {\bibfnamefont {J.~M.}\ \bibnamefont {Tranquada}}, \bibinfo {author}
  {\bibfnamefont {S.}~\bibnamefont {Wakimoto}}, \bibinfo {author}
  {\bibfnamefont {G.}~\bibnamefont {Xu}}, \ and\ \bibinfo {author}
  {\bibfnamefont {K.}~\bibnamefont {Yamada}},\ }\href {\doibase
  10.1143/JPSJ.81.011007} {\bibfield  {journal} {\bibinfo  {journal} {Journal
  of the Physical Society of Japan}\ }\textbf {\bibinfo {volume} {81}},\
  \bibinfo {pages} {011007} (\bibinfo {year} {2012})}\BibitemShut {NoStop}%
\bibitem [{\citenamefont {Vignolle}\ \emph {et~al.}(2007)\citenamefont
  {Vignolle}, \citenamefont {Hayden}, \citenamefont {McMorrow}, \citenamefont
  {Ronnow}, \citenamefont {Lake}, \citenamefont {Frost},\ and\ \citenamefont
  {Perring}}]{hourglass2}%
  \BibitemOpen
  \bibfield  {author} {\bibinfo {author} {\bibfnamefont {B.}~\bibnamefont
  {Vignolle}}, \bibinfo {author} {\bibfnamefont {S.~M.}\ \bibnamefont
  {Hayden}}, \bibinfo {author} {\bibfnamefont {D.~F.}\ \bibnamefont
  {McMorrow}}, \bibinfo {author} {\bibfnamefont {H.~M.}\ \bibnamefont
  {Ronnow}}, \bibinfo {author} {\bibfnamefont {B.}~\bibnamefont {Lake}},
  \bibinfo {author} {\bibfnamefont {C.~D.}\ \bibnamefont {Frost}}, \ and\
  \bibinfo {author} {\bibfnamefont {T.~G.}\ \bibnamefont {Perring}},\ }\href
  {\doibase 10.1038/nphys546} {\bibfield  {journal} {\bibinfo  {journal} {Nat.
  Phys.}\ }\textbf {\bibinfo {volume} {3}},\ \bibinfo {pages} {163} (\bibinfo
  {year} {2007})}\BibitemShut {NoStop}%
\bibitem [{\citenamefont {Cheong}\ \emph {et~al.}(1991)\citenamefont {Cheong},
  \citenamefont {Aeppli}, \citenamefont {Mason}, \citenamefont {Mook},
  \citenamefont {Hayden}, \citenamefont {Canfield}, \citenamefont {Fisk},
  \citenamefont {Clausen},\ and\ \citenamefont
  {Martinez}}]{PhysRevLett.67.1791}%
  \BibitemOpen
  \bibfield  {author} {\bibinfo {author} {\bibfnamefont {S.-W.}\ \bibnamefont
  {Cheong}}, \bibinfo {author} {\bibfnamefont {G.}~\bibnamefont {Aeppli}},
  \bibinfo {author} {\bibfnamefont {T.~E.}\ \bibnamefont {Mason}}, \bibinfo
  {author} {\bibfnamefont {H.}~\bibnamefont {Mook}}, \bibinfo {author}
  {\bibfnamefont {S.~M.}\ \bibnamefont {Hayden}}, \bibinfo {author}
  {\bibfnamefont {P.~C.}\ \bibnamefont {Canfield}}, \bibinfo {author}
  {\bibfnamefont {Z.}~\bibnamefont {Fisk}}, \bibinfo {author} {\bibfnamefont
  {K.~N.}\ \bibnamefont {Clausen}}, \ and\ \bibinfo {author} {\bibfnamefont
  {J.~L.}\ \bibnamefont {Martinez}},\ }\href {\doibase
  10.1103/PhysRevLett.67.1791} {\bibfield  {journal} {\bibinfo  {journal}
  {Phys. Rev. Lett.}\ }\textbf {\bibinfo {volume} {67}},\ \bibinfo {pages}
  {1791} (\bibinfo {year} {1991})}\BibitemShut {NoStop}%
\bibitem [{\citenamefont {Kim}\ \emph {et~al.}(2014{\natexlab{b}})\citenamefont
  {Kim}, \citenamefont {Daghofer}, \citenamefont {Said}, \citenamefont {Gog},
  \citenamefont {Brink}, \citenamefont {Khaliullin},\ and\ \citenamefont
  {Kim}}]{kim2014excitonic}%
  \BibitemOpen
  \bibfield  {author} {\bibinfo {author} {\bibfnamefont {J.}~\bibnamefont
  {Kim}}, \bibinfo {author} {\bibfnamefont {M.}~\bibnamefont {Daghofer}},
  \bibinfo {author} {\bibfnamefont {A.~H.}\ \bibnamefont {Said}}, \bibinfo
  {author} {\bibfnamefont {T.}~\bibnamefont {Gog}}, \bibinfo {author}
  {\bibfnamefont {J.~v.~d.}\ \bibnamefont {Brink}}, \bibinfo {author}
  {\bibfnamefont {G.}~\bibnamefont {Khaliullin}}, \ and\ \bibinfo {author}
  {\bibfnamefont {B.~J.}\ \bibnamefont {Kim}},\ }\href {\doibase
  10.1038/ncomms5453} {\bibfield  {journal} {\bibinfo  {journal} {Nat.
  Commun.}\ }\textbf {\bibinfo {volume} {5}},\ \bibinfo {pages} {4453}
  (\bibinfo {year} {2014}{\natexlab{b}})}\BibitemShut {NoStop}%
\bibitem [{\citenamefont {Keimer}\ \emph {et~al.}(1993)\citenamefont {Keimer},
  \citenamefont {Birgeneau}, \citenamefont {Cassanho}, \citenamefont {Endoh},
  \citenamefont {Greven}, \citenamefont {Kastner},\ and\ \citenamefont
  {Shirane}}]{Keimer1993}%
  \BibitemOpen
  \bibfield  {author} {\bibinfo {author} {\bibfnamefont {B.}~\bibnamefont
  {Keimer}}, \bibinfo {author} {\bibfnamefont {R.~J.}\ \bibnamefont
  {Birgeneau}}, \bibinfo {author} {\bibfnamefont {A.}~\bibnamefont {Cassanho}},
  \bibinfo {author} {\bibfnamefont {Y.}~\bibnamefont {Endoh}}, \bibinfo
  {author} {\bibfnamefont {M.}~\bibnamefont {Greven}}, \bibinfo {author}
  {\bibfnamefont {M.~A.}\ \bibnamefont {Kastner}}, \ and\ \bibinfo {author}
  {\bibfnamefont {G.}~\bibnamefont {Shirane}},\ }\href {\doibase
  10.1007/BF01344067} {\bibfield  {journal} {\bibinfo  {journal} {Z. Phys. B}\
  }\textbf {\bibinfo {volume} {91}},\ \bibinfo {pages} {373} (\bibinfo {year}
  {1993})}\BibitemShut {NoStop}%
\bibitem [{\citenamefont {Ye}\ \emph {et~al.}(2013)\citenamefont {Ye},
  \citenamefont {Chi}, \citenamefont {Chakoumakos}, \citenamefont
  {Fernandez-Baca}, \citenamefont {Qi},\ and\ \citenamefont
  {Cao}}]{PhysRevB.87.140406}%
  \BibitemOpen
  \bibfield  {author} {\bibinfo {author} {\bibfnamefont {F.}~\bibnamefont
  {Ye}}, \bibinfo {author} {\bibfnamefont {S.}~\bibnamefont {Chi}}, \bibinfo
  {author} {\bibfnamefont {B.~C.}\ \bibnamefont {Chakoumakos}}, \bibinfo
  {author} {\bibfnamefont {J.~A.}\ \bibnamefont {Fernandez-Baca}}, \bibinfo
  {author} {\bibfnamefont {T.}~\bibnamefont {Qi}}, \ and\ \bibinfo {author}
  {\bibfnamefont {G.}~\bibnamefont {Cao}},\ }\href {\doibase
  10.1103/PhysRevB.87.140406} {\bibfield  {journal} {\bibinfo  {journal} {Phys.
  Rev. B}\ }\textbf {\bibinfo {volume} {87}},\ \bibinfo {pages} {140406}
  (\bibinfo {year} {2013})}\BibitemShut {NoStop}%
\bibitem [{\citenamefont {Dhital}\ \emph {et~al.}(2013)\citenamefont {Dhital},
  \citenamefont {Hogan}, \citenamefont {Yamani}, \citenamefont {de~la Cruz},
  \citenamefont {Chen}, \citenamefont {Khadka}, \citenamefont {Ren},\ and\
  \citenamefont {Wilson}}]{PhysRevB.87.144405}%
  \BibitemOpen
  \bibfield  {author} {\bibinfo {author} {\bibfnamefont {C.}~\bibnamefont
  {Dhital}}, \bibinfo {author} {\bibfnamefont {T.}~\bibnamefont {Hogan}},
  \bibinfo {author} {\bibfnamefont {Z.}~\bibnamefont {Yamani}}, \bibinfo
  {author} {\bibfnamefont {C.}~\bibnamefont {de~la Cruz}}, \bibinfo {author}
  {\bibfnamefont {X.}~\bibnamefont {Chen}}, \bibinfo {author} {\bibfnamefont
  {S.}~\bibnamefont {Khadka}}, \bibinfo {author} {\bibfnamefont
  {Z.}~\bibnamefont {Ren}}, \ and\ \bibinfo {author} {\bibfnamefont {S.~D.}\
  \bibnamefont {Wilson}},\ }\href {\doibase 10.1103/PhysRevB.87.144405}
  {\bibfield  {journal} {\bibinfo  {journal} {Phys. Rev. B}\ }\textbf {\bibinfo
  {volume} {87}},\ \bibinfo {pages} {144405} (\bibinfo {year}
  {2013})}\BibitemShut {NoStop}%
\bibitem [{\citenamefont {Solovyev}\ \emph {et~al.}(2015)\citenamefont
  {Solovyev}, \citenamefont {Mazurenko},\ and\ \citenamefont
  {Katanin}}]{PhysRevB.92.235109}%
  \BibitemOpen
  \bibfield  {author} {\bibinfo {author} {\bibfnamefont {I.~V.}\ \bibnamefont
  {Solovyev}}, \bibinfo {author} {\bibfnamefont {V.~V.}\ \bibnamefont
  {Mazurenko}}, \ and\ \bibinfo {author} {\bibfnamefont {A.~A.}\ \bibnamefont
  {Katanin}},\ }\href {\doibase 10.1103/PhysRevB.92.235109} {\bibfield
  {journal} {\bibinfo  {journal} {Phys. Rev. B}\ }\textbf {\bibinfo {volume}
  {92}},\ \bibinfo {pages} {235109} (\bibinfo {year} {2015})}\BibitemShut
  {NoStop}%
\end{thebibliography}%

\end{document}